\documentclass[useAMS,usenatbib,referee]{mn2e}

\usepackage{graphicx}
\title[Mass Loss in Dwarf Galaxies]
{The mass loss process in dwarf galaxies from 3D hydrodynamical simulations: the 
role of dark matter and starbursts}\author[L. O. Ruiz, D. Falceta-Gon\c{c}alves, 
G. A. Lanfranchi \& A. Caproni]{L. O. Ruiz$^{1}$, 
D. Falceta-Gon\c{c}alves$^{2}$\thanks{E-mail:dfalceta@usp.br}, G. A. Lanfranchi$^{1}$ \& A. Caproni$^{1}$ 
\\$^{1}$N\' ucleo de Astrof\'
isica Te\' orica, Universidade Cruzeiro do Sul - Rua Galv\~ ao Bueno
868, CEP 01506-000, S\~ao Paulo, Brazil, \\ $^{2}$EACH, 
Universidade de S\~ao Paulo, Rua Arlindo Bettio 1000, CEP 03828-000,
S\~ao Paulo, Brazil}

\begin{document}

\date{}

\pagerange{\pageref{firstpage}--\pageref{lastpage}} \pubyear{2012}

\maketitle

\label{firstpage}

\begin{abstract} 
 
Theoretical $\Lambda$CDM cosmological models predict a much larger number of low mass 
dark matter haloes than has been observed in the Local Group of galaxies. One possible 
explanation is the increased difficulty of detecting these haloes if most of the 
visible matter is lost at early evolutionary phases through galactic winds. In this work 
we study the current models of triggering galactic winds in dwarf spheroidal galaxies 
(dSph) from supernovae, and study, based on 3D hydrodynamic numerical simulations, 
the correlation of the mass loss rates and important physical parameters 
as the dark matter halo mass and its radial profile, and the star formation rate. We find that the existence of winds is ubiquitous, independent on the gravitational potential. 
Our simulations revealed that the Rayleigh-Taylor Instability (RTI) may play 
a major role on pushing matter out of these systems, even for very massive haloes. The instability is responsible for $5 - 40$\% of the mass loss during the early evolution of the galaxy, being less relevant at $t > 200$Myrs.  There is no significant difference in the mass loss rates obtained for the different dark matter profiles studied (NFW and logarithmic). 
We have also found a correlation between the mass loss rate 
and both the halo mass and the rate of supernovae, as already reported in previous works. 
Besides, the epoch in which most of the baryon galactic matter is removed from the galaxy varies depending on the SN rate and gravitational potential. The later, combined to the importance of the RTI in each model, may change our understanding about the chemical evolution of dwarf galaxies, as well as in the heavy element contamination 
of the intergalactic medium at high redshifts.

\end{abstract}

      
\section{Introduction}

Tens of dwarf spheroidal galaxies (dSph, hereafter) are known to populate 
the Local Group of galaxies (Mateo 1998), including the so-called classical dSph (Sculptor, Fornax, Ursa Minor, Leo I, Leo II, Draco, Sextans, Carina, and Sagittarius) and the recently discovered ones (normally called ultrafaint dwarfs), observed through analysis of the deep multi-colour photometry from the Sloan Digital Sky Survey (SDSS) (Willman et al. 2005A, 2005B,  Zucker et al. 2006A, 2006B, Belokurov et al. 2006, Belokurov et al. 2007, Irwin et al. 2007, Walsh et al. 2007, Belokurov et al. 2008, Belokurov et al. 2009, Grillmair 2009, Watkins et al. 2009, Belokurov et al. 2010). 

The classical dSph galaxies are relatively large in size, typically with half light radii within the range of $100 - 500$pc, but present very low luminosity, $M_V > -11$ (Mateo 1998). Their sizes are related to the large velocity dispersion of the stars, which indicates the presence of large amounts of dark matter (Armandroff, Olszewski \& Pryor 1995, Mu\~noz et al. 2005, 2006, Walker 2012 and references therein). The dSph galaxies are therefore considered dark matter dominated systems, with mass-to-luminosity ratios $M/L > 20$. The ultrafaint dwarfs, on the other hand, seem to exhibit even larger mass-to-luminosity ratios, with values varying from hundreds (Kleyna et al. 2005, Mu�oz et al. 2006, Martin et al. 2007, Simon \& Geha 2007) up to $M/L_V \sim 3400$, as observed in Segue 1 (Geha et al. 2009, Simon et al. 2011). Consequently, both classical and ultrafaint dSph galaxies must be related to the dark matter halos associated to our Galaxy as predicted by the Cold Dark Matter (CMD) cosmological simulations (Moore et al. 1999). However, even with the recent discoveries of faint objects, the number of dSph associated with the Milky Way ($\sim$ 26 objects) is much lower than the theoretical predictions (of $\sim$ hundreds of objects) from the current cosmological paradigm of the cold dark matter ($\Lambda$CDM) Universe (Moore et al. 1999, Kyplin et al. 1999). 

A proposed scenario considers a threshold in the
dark matter mass of the primordial haloes which would host a dwarf galaxy, to reconcile the discrepancy between the predictions of cosmological simulations and observations. The limiting mass is defined according to the capability of the galaxy to retain gas and maintain star formation despite the destructive processes during early star formation, such as supernovae (SNe) explosions and heating of the interstellar medium (ISM) by photoionizing radiation (Somerville 2002, Benson et al. 2002, Hayashi et al. 2003, Kazantzidis et al. 2004).
With the mass threshold the number of existing dwarf galaxies would be considerably lower, close to what is observed (Kazantzidis et al. 2004). However, the state-of-art cosmological simulations are able to describe the dynamics of the dark matter component only, being unable to properly describe the contribution and effects of baryons (Diemand et al. 2007, Springel et al. 2008, Boylan-Kolchin et al. 2012). It has been shown by several high resolution simulations, for instance, that star formation and supernovae feedback can alter the distribution and the profile of the dark matter component of the galaxy (Read \& Gilmore 2005, Mashchenko et al. 2006, Governato et al. 2010, Cloet-Osselaer et al. 2012, Pontzen \& Governato 2012, Governato et al. 2012). A cuspy dark matter density profile can be turned into a cored profile after the beginning of the star formation (Pasetto et al. 2010, Governato et al. 2012).

When baryons are not included, coalescence of dark matter haloes, mergers and tidal interactions with more massive galaxies could, in principle, explain the increase in the mass-to-luminosity ratio of these objects. Pe\~narrubia et al. 
(2008) studied the halo masses of the Local Group dSphs based on the 
velocity dispersion of the luminous component. Fitting Navarro-Frenk-White (Navarro et al. 1996, NFW hereafter) profiles, 
virial masses of $M_{\rm V} \sim 10^9 $M$_{\odot}$ at virial radii $R_{\rm V} 
\sim 1.5-7.0$kpc, with core masses $M_{\rm c} \sim 10^7 $M$_{\odot}$ 
at core radii $R_{\rm c} \sim 0.2-0.4$kpc, are found. Besides, Pe\~narrubia et al. 
(2008) state that the velocity dispersion of the luminous component peaks at radii 
much lower than the virial radius (see also the review of Walker 2012). Therefore, stars are usually packed in a region small enough to be unaffected by tidal disruption produced, for instace, by gravitational interaction due to neighboring galaxies.

Another possibility would be that these low luminosity galaxies are 
the result of quenched star formation, if the effects of baryons are accounted. Large feedback from starbursts would work on two ways, firstly on radiative suppression of star formation (e.g. Andersen \& Burkert 2000) and secondly on the ejection of gas in form of galactic winds (e.g. Dekel \& Silk 1986, Somerville \& Primack 1999, Ferrara \& Tolstoy 2000). Actually, radiative suppression only retards the star formation to the cooling timescales. 

The idea of galactic mass-loss from starbursts in galaxies is not new 
(Matthews \& Baker 1971, Larson 1974, Bradamante, Matteucci \& D`Ercole 
1998, Mac Low \& Ferrara 1999, Fragile et al. 2003, Sawala et al. 2009, Stringer et al. 2011, Zolotov et al 2012, and many others). Despite of the great theoretical development of this 
field, open issues still remain as the results strongly depend on the 
modeling itself. As an example, star formation efficiencies $\leq 10$\% 
would be enough to processed gas to leave the gravitational potential 
(Mori, Ferrara, \& Madau 2002), but this conclusion was obtained 
considering a single and localized burst of SNe. Fragile et al. (2003)
studied the chemical enrichment in dSphs considering the ejection of 
processed material by SNe and found that a SNe rate of $10^{-6}$yr$^{-1}$ 
is enough to remove half of the baryonic matter out of these galaxies, 
though in their paper cooling effects are treated inhomogeneously within
the computed volume to account for a less efficient cooling in the SNe. 
The effects of mass-loss processes in the evolution of dSphs can also be noticed in the chemical enrichment of such galaxies (Marcolini et al. 2006, 2008, Lanfranchi \& Matteucci 2007, Salvadori, Ferrara \& Schneider 2008, Revaz et al. 2009, Okamoto et al. 2010, Sawala et al. 2010, Kirby, Martin \& Finlator 2011, Revaz \& Jablonka 2012, and many others).

Bradamante, Matteucci \& D'Ercole (1999) proposed a semi-analytic chemical evolution model for blue compact galaxies (BCG) and dwarf irregular galaxies (DIG) based on a detailed calculation of winds generated from SNe feedback. In order to explain the observed abundances 
of elements the winds should be able to completely remove the ISM gas 
out of the galaxies, mostly due to type II SNe. Also, an unrealistic 
amount of dark matter would be necessary to avoid this process to occur. 
Qualitative similar results were obtained by Recchi, Matteucci \& D'Ercole 
(2001) based on 2D hydrodynamical simulations, with additional conclusions 
that mostly heavy elements are dragged out of the galaxies by the 
galactic winds. Kirby, Martin \& Finlator (2011) argued that galactic winds affect substantially the metallicity of Milky Way satellites, since their low observed values are systematically more metal-poor than expected from ``closed-box" chemical evolution models. Lanfranchi \& Matteucci (2007) by adopting a chemical evolution model compared to observed stellar metallicity distributions in Ursa Minor and Draco, claimed that strong galactic winds are required to repoduce the observed data. Using hydrodynamical simulations, Marcolini et al. (2006, 2008) predicted weak winds driven by supernovae explosions and suggested the need of an external mechanism (such as ram pressure or tidal stripping) to account for the low fraction of gas in dSph galaxies. Revaz et al. (2009) and Revaz \& Jablonka (2012) reached similar conclusions through a SPH code: the chemical properties of local dSph galaxies can be achieved only if a high rate of gas loss is considered. Salvadori, Ferrara \& Schneider (2008) with a semi-analytical code also concluded that a large fraction of gas can be removed from dSph galaxies as a result of star formation feedback. Therefore, the mass-loss process during the early stages of dSph galaxy formation is fundamental for the understanding of galaxy formation as a whole and for comparison of observations and cosmological simulations of structure formation. 
This, however, depends on both the star formation rates at early stages and the dark matter gravitational potential. These studies however diverge on the precise mass-loss rates obtained, basically due to the different prescription of SNe driven wind used. The timescales needed for the gas removal is also a matter of debate. Also, the models still cannot properly predict if these galactic winds are more efficient in removing enriched or the low metallicity portion of the interstellar gas. The current techniques employed can lead to conclusions quite different. Also, the role of different dark matter distributions in the gas removal efficiency has not been addressed yet.

In this work we focus on modelling the dynamics of the baryonic content of a typical isolated dSph galaxy, and study the mass loss process in this type of galaxy based on hydrodynamical numerical simulations that follow the evolution of the mass content of the system. We developed a numerical setup for 
typical dSphs, taking into account the gravitational potential of the dark matter content and the star formation activity. The later, in particular, is treated in a more consistent way as done in previous works. The theoretical basis of the model and the numerical setup are described in Section 2. In Section 
3 we present the results, followed by a discussion of the main results and comparison to previous works. The main conclusions of this work are presented in Section 4.

\section{The model}

In this work we simulate the dynamical evolution of the gas component of a typical dSph galaxy taking into account the energy feedback from SNe, the gravitational potential of a given stationary dark matter distribution, and thermal losses of the heated gas, as described below.

\subsection{Numerical Setup}

The simulations were performed solving the set of hydrodynamical equations, in 
conservative form, as follows:

\begin{equation}
\frac{\partial \rho}{\partial t} + \mathbf{\nabla} \cdot (\rho{\bf v}) = 0,
\end{equation}

\begin{equation}
\frac{\partial \rho {\bf v}}{\partial t} + \mathbf{\nabla} \cdot \left[ \rho{\bf v v} + 
p {\bf I} \right] = {\bf f},
\end{equation}

\noindent
where $\rho$, ${\bf v}$ and $p$ are the plasma density, velocity and pressure, 
respectively, and ${\bf f}$ represents the external 
source terms (e.g. the dark matter gravity). 

The set of equations is complete with an explicit 
equation for the evolution of the energy. 
For the radiative cooling the set of equations is closed calculating 

\begin{equation}
\frac{\partial P}{\partial t} = \frac{1}{(1-\gamma)} n^2 \Lambda(T),
\end{equation}

\noindent
after each timestep, where $n$ is the number density, $P$ is the gas pressure, $\gamma$ 
the adiabatic constant and $\Lambda(T)$ is the interpolation function from an electron 
cooling efficiency table for an optically thin gas (see Falceta-Gon\c calves et al. 2010a, 
for details). 

The equations are solved using a second-order shock capturing
Godunov-type scheme, with essentially nonoscillatory
spatial reconstruction. The time integration is performed with a Runge\-Kutta (RK) method. We used 
HLLC Riemann solver (see Mignone 2007) to obtain the
numerical fluxes at each step to properly resolve the contact discontinuities and reduce the
numerical difusion at the shocks. The code has been tested extensively in previous work as tool to study several different astrophysical problems (Kowal et al. 2009, Le\~ao et al 2009, Burkhart et al. 2009, Falceta-Gon\c calves et al. 2010a,b, Falceta-Gon\c calves, Lazarian \& Houde 2010, Kowal, Falceta-Gon\c calves \& Lazarian 2011, Falceta-Gon\c calves \& Lazarian 2011, Monteiro \& Falceta-Gon\c calves 2011, Falceta-Gon\c calves \& Abraham 2012).

The code is run in the single fluid approximation and is not able to solve separately the cooling 
functions for the initially set ISM (with low metallicity) and for the the metal enriched gas ejected by the SNe. For the sake of simplicity we used a low metallicity abundance $Z=10^{-3} Z_{\odot}$ for the cooling function in all regions of the computational domain, during the whole computational time.  We also perfomed 2-dimensional tests comparing the evolution of the gas for $Z=10^{-3} Z_{\odot}$ and $Z=Z_{\odot}$, and found no significant differences. The major different in cooling rates are given at $\sim 10^5$K, where $C$ and $O$ emission lines dominate the cooling rate for $Z=Z_{\odot}$, being approximately one order of magnitude larger than the cooling rate for $Z=10^{-3} Z_{\odot}$. In both cases though the cooling timescales are typically similar, when compared to the dynamical timescales.

The gas is initially set in isothermal hydrostatic equilibrium with the dark matter gravitational potential, where we assumed $T_{\rm gas}=10^4$K. The density profile peaks at the center of the box with an assumed density $n_c^{\rm gas} = 1$ or 10 cm$^{-3}$, depending on the model (e.g. Mori, Ferrara \& Mori 2002) , resulting in a total initial gas mass of $M_{\rm gas} \simeq 10^6$M$_{\odot}$ and $10^7$M$_{\odot}$, respectively. The gas to dark matter mass ratio, calculated at the virial radius, is therefore in the range of $0.001 - 0.01$, but much larger at the half light radii.

We evolve the dynamics of the galactic gas in a 3-dimensional domain of $512^3$ cells, equally spaced in an uniform grid, totaling a physical lenght of 1kpc in each cartesian direction. The grid resolution in physical units is equivalent to $\sim 1.95$pc/cell. The simulations were run up to $t=1$Gyr.

\subsubsection{The injection of SNe energy}

The injection of energy in the code should mimic the effects of real SNe explosions in the ISM. In our model, at a given time $t$, the density distribution of gas is analyzed and all cells with densities larger than the threshold of $0.1$cm$^{-3}$ are selected as possible sites of starburst. The ad hoc assumption of a low average density limit for star formation is due to the numerical resolution of our simulations. It is still not possible to completely follow the turbulent evolution of the ISM, the creation of parsec scale dense clouds, and their fragmentation to sub-parsec scales. 

The second step is to randomly choose the cells that will have a starburst at that timestep, weighted linearly by the local density, i.e. denser regions have larger probability as given by the Schmidt-Kennicut law. The total number of cells chosen per timestep to present a starburst is determined by the preset star formation rate of the galaxy. It is worth to mention that there is no special need for defining a specific range of temperatures in which star formation would occur. This is basically due to the fact that, if the density is large, the typical motions of the diffuse medium always end up with the creation of a thermally unstable region that will develop into a molecular cloud, or fragment into many. 

Each selected cell is tagged and we inject thermal energy, proportional to the number of SNe expected to occur based on the SK law for the local density (see Equations 5 and 8 in the next section, but with $\eta = 1$ and $\epsilon_{51} = 1$). Notice that in this prescription the injection of energy is not fixed spacially, nor in time, i.e. there is no fixed star formation rate for the galaxy. Depending on the dark matter mass there will be more or less concentration of gas in the core, which will naturally result in a larger concentration of SNe in this region. Also, instead of injecting hundreds of SNe at once we allow partial evolution of the gas from the first bursts before all energy is released from all SNe. Finally, another advantage of this type of modeling is that if too much gas is removed from the galaxy the star formation is naturally quenched, instead of injecting energy even when voids are created in the simulated ISM.

\subsection{The physics of the galactic mass-loss}

Larson (1974) had already proposed a quantitative model 
to show that a fraction of the SNe energy would be converted into kinetic 
and thermal pressures, which result in galactic winds. For example, if 
radiative cooling is not efficient, a simple energy conservation approach 
may be used, i.e. the energy from SNe is converted into the ISM gas thermal 
and turbulent kinetic energies. These may be compared to the binding energy 
of the gravitational potential, as follows (Bradamante, Matteucci \& D'Ercole 1998):

\begin{equation}
E_{\rm kin,th} (t) \ge E_b (t),
\end{equation}

\noindent
where

\begin{equation}
E_{\rm kin,th} (t) = \int_{0}^{t}{\eta \epsilon_{\rm SN} R_{\rm SN}(t')dt'},
\end{equation}

\noindent
being $\eta$ the efficiency in the conversion of the SN energy into kinetic-thermal pressure of the gas\footnote{typical efficiencies of $\sim 1 - 5\%$ have been used in both numerical and semi-analytical models (see Bradamante, Matteucci \& D'Ercole 1998). In this work, however, we do not set an {\it ad hoc} thermal to kinetic energy efficiency since our numerical prescription follows the thermal evolution of the SN burst self-consistently. Notice that we obtain efficiencies as low as 1\%, and as high as 60\%, depending on the surrounding density and the evolution of the buoyancy instability. The main issue in modelling the SN-driven wind is not the value of $\eta$, but its dependence on the local physical parameters.}, 
$\epsilon_{\rm SN}$ the average energy released by a supernova and $R_{\rm SN}$ the SN occurrence rate, given by:

\begin{equation}
R_{\rm SN} (t) \simeq \int_{8}^{120}{sSFR (t-\tau_M) \phi (M) dM},
\end{equation}

\noindent
where $sSFR$ is the specific star formation rate, i.e. the star formation rate divided by the total mass of 
stars, $\tau_M \sim 1.2 M^{-1.85}+0.003$ Gyrs is the evolution lifetime of a star with a given mass $M$ 
(Matteucci \& Greggio 1986), and $\phi$ is the adopted initial mass function (IMF). 

Notice that only SNe of type II were taken into account in $R_{\rm SN}$, as calculated above. Matteucci \& 
Greggio (1986) calculated the supernova occurrence rates for both types Ia and II, showing that SNe type II are 
dominant by an order of magnitude at $t < 1$Gyr, and continues more than a factor of 2 larger even for $t \gg 1$Gyr. 
Obviously, even though basically irrelevant for the energy budget of a galaxy, the type Ia SNe rate must be 
accounted for the chemical evolution models since most of the Fe group elements of the ISM is originated in these 
objects. 

Therefore, if only the massive stars ($M_* > 8$M$_{\odot}$) are considered the IMF is reduced to the single-slope 
form $\phi (M) \sim 0.17 M^{-2.3}$ (Kroupa 2001). Finally, in order to obtain $R_{\rm SN}$ a specific SFR function is needed. 

For the case of a delta function SFR in time, i.e. a single and very fast starburst, the SN rate ($R_{\rm SN}$) can be written as:

\begin{equation}
R_{\rm SN} (t) \propto \int_{8}^{120}{\delta (t-\tau_M) M^{-2.3} dM},
\end{equation}

\noindent
which, by changing variables $M$ to $\tau$ is reduced to:

\begin{eqnarray}
R_{\rm SN} (t) \propto \int{\delta (t-\tau_M) (\tau - 0.003)^{1.0175} d\tau} \nonumber \\
 \propto (t_{\rm Gyr}-0.003)^{1.0175}.
\end{eqnarray}

\noindent
The rate of type II SNe for a single starburst increases with time, with a peak at $\sim 20$Myrs, as the number of very massive stars that explode at short timescales is much smaller than of those that explode at later times.

In reality, the star formation rate is a function of time. The Schmidt-Kennicutt empirical law parameterizes the 
SFR with both local and global properties of the system: the local density and a dynamical timescale, respectively, 
i.e (Schmidt 1956, Kennicutt 1998):

\begin{equation}
SFR \propto \frac{\rho^k}{\tau_{\rm dyn}},
\end{equation}

\noindent
being $k\sim 1.4$.
For spiral galaxies, such as the Milky Way, this empirical law has the volume gas density $\rho$ replaced by 
$\Sigma$, the surface gas density, and $\tau_{\rm dyn}$ by the typical rotation period around the center 
of the galaxy $\Omega$. 
In a more realistic scenario, in which the SFR is a function of time, the peak will occur at different times. 
In Figure 1 we show the dependency of the SNR with time for an exponentially decaying SFR (with $\tau$ being 
the scale of the decay).

\begin{figure}
	\centering
		\includegraphics[scale=0.35]{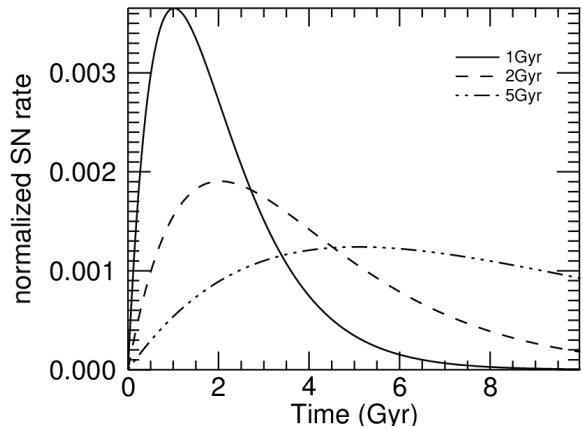}
		\caption {Type II supernova occurence rate for a single starburst with an exponentially decaying star formation rate, i.e. $SFR \propto exp(-t/\tau)$. Three SFR decay timescales are considered, $\tau = 1$(solid), 2 (dashed) and 5Gyrs (dash-dotted). Each curve is normalized to its integral over time.}
	\label{snrate}
\end{figure}

Basically, if no energy losses are considered and knowing the star formation history, Eq.2 may be computed at a 
given time $t$ and directly compared to the gravitational binding energy. In a real system however it is not 
straightforward. This because the kinetic energy is not simply cumulative for large timescales, 
i.e. the time evolution of $E_{\rm kin,th}$ is very complex, and many processes may take place, such as 
turbulent dissipation, viscosity, and the radiative cooling. Finally, kinetic energy is also lost when 
part of the gas is removed through the winds. This loss is generally disregarded in semi-analytical models. 
The absence of dissipation effects results in an underestimation of the minimum 
SNe rate needed to remove the gas from galaxies.

\subsubsection{The role of radiative cooling}

The efficiency of SNe in pushing matter out of galaxies depends strongly on the radiative cooling. During the 
early expansion stage of a SN blast the gas temperature of the ejecta is very high ($>10^6$K), and its expansion 
speed above $v_{\rm SN}>3000$km s$^{-1}$. Therefore, the expansion timescale, $t_{\rm dyn} \sim R/v$, being $R$ 
the radius of the shell, is much shorter than the cooling timescale $t_{\rm cool} \sim kT/n\Lambda$, where 
$\Lambda$ stands for the cooling function. As the shell expands the temperature decreases and the velocity of 
the ejecta diminishes in a sense that the cooling becomes faster than the dynamical timescales. At this stage, 
which occurs after $t_{\rm c} \sim 10^4 - 10^5 \epsilon_{51}^{4/17} n^{-9/17}$yr (Cox 1972, Le\~ao et al. 2009), 
where $\epsilon_{51}$ represents the energy released by the SN in units of $10^{51}$ergs and $n$ is the ISM gas 
density in $cm^{-3}$, the typical radius of the shell is $R_{\rm c} \sim 20 - 30 \epsilon_{51}^{5/17} n^{-7/17}$pc.
Further expansion of the shell will be greatly reduced due to the conversion of the kinetic energy of the blast 
wave into radiation. 

The typical sizes of the SNe shells are, in general, much smaller than the galaxies radii. Considering then that 
most of the SNe occur near the center of the galaxies multiple explosions in a short timescale are required in 
order to generate a galactic wind. 

A realistic description of the evolution of SNe ejecta, and their interaction with the ISM, is too 
complex to be handled analytically. Nonlinear evolution of bubbles, turbulence and presence multiple asymmetric 
shocks, at different stages of evolution are a few examples of the physics that must be taken into account in 
this problem. For this reason, full 3D numerical simulations are required. 

\subsection{Dark matter density profiles}

Not only the total mass but also the density profiles of the dark matter haloes have an important role on the physics of mass loss in dSphs (Falceta-Gon\c calves 2012). The observational detection of the dark matter haloes in dwarf galaxies is not an easy task. Their intrinsic low luminosity, absence of diffuse X-ray emission and history of tidal interactions makes it even more complicated. From theoretical point of view, CDM cosmological simulations predict the existence of cusped profiles (e.g., Navarro et al. 1996), which means the dark matter density diverges formally towards the centre of galaxies.  The NFW dark matter profile is perhaps the most famous member of cusped dark matter distributions, which is defined as:

\begin{equation}
\rho_{\rm DM}\left( r \right)=\frac{200}{3} \frac{A\left( c \right) \rho_{\rm crit}}{\left( r/r_{\rm s} \right) 
\left( 1+r/r_{\rm s} \right)^2}
\end{equation}

\noindent
where 

\begin{equation}
A\left( c \right) = \frac{c^3}{\ln{\left(1+c\right)-c/\left(1+c\right)}},
\end{equation}

\noindent
$\rho_{\rm crit}$ is the critical density for a closed Universe, $c$ a concentration parameter and $r_{\rm s}$ 
the distribution length scale.

 However, recent observational and numerical works have shown that cusped dark matter profiles do not fit well the observational line-of-sight velocity dispersions of a large number of dSph galaxies, as well as many other types of dwarf galaxies, such as the low surface brightness (LSB) disk galaxies, where the gravitational potential at the central regions of the galaxies tends to be less steep and cored (e.g. Burkert 1995; van den Bosch et al. 2000; de Blok \& Bosma 2002; Kleyna et al. 2003; Simon et al. 2005; Walker et al. 2009; Governato et al. 2010; Oh et al. 2011; Del Popolo 2012; Jardel \& Gebhardt 2012). Many core dark matter profiles have been suggested in the literature to account for the dark matter distribution in dwarf galaxies (e.g., Begeman  1989; Begeman, Broeils \& Sanders 1991; Burkert 1995; de Blok et al. 2001; Blais-Ouellette et al. 2001; Simon et al. 2005; Amorisco \& Evans 2012; Zolotov et al 2012). A representative of this type of density profile can be derived from a logarithmic potential, as follows:

\begin{equation}
\Phi\left( r \right) \simeq \frac{V_{\rm c}^2}{2} \ln{(r^2+r_{\rm c}^2)},
\end{equation}

\noindent
where $r_{\rm c}$ is the core radius and $V_{\rm c}$ is the circular speed at $r \rightarrow \infty$, result 
in density distributions as follows:

\begin{equation}
\rho_{\rm DM}\left( r \right) \simeq \frac{V_{\rm c}^2}{4\pi G} \frac{3r_{\rm c}^2+r^2}{\left(r^2+r_{\rm c}^2\right)^2},
\end{equation}

\noindent
which features a flat core and is $\propto r^{-2}$ for $r \gg r_{\rm c}$, similar to the observations (e.g. 
Jardel \& Gebhardt 2012). 

The inferred discrepancy between the central dark matter distributions in dwarf galaxies and those predicted from the CDM numerical simulations is known as the �cusp/core� problem. Time evolution from an initial cusped distribution (as predicted from the CDM simulations) to a core-like density profile induced by baryonic feedback from supernovae winds (e.g., Governato et al. 2010; de Souza et al. 2011; Pontzen \& Governato 2012; Governato et al. 2012), and/or tidal stirring of rotationally supported dwarf galaxies (e.g, Mayer et al. 2001a, 2001b; Klimentowski et al. 2007, 2009; Kazantzidis et al. 2011; Lokas, Karantzidis \& Mayer 2012) seem to be the most natural candidates for solving such discrepancy.

The main mechanisms that result in the ejection of matter out of the galaxies operate mainly at within the galactic cores, i.e. even for haloes with similar masses, the radial distribution of dark matter plays a role on the resulting mass loss rates. It is important to emphasize that this aspect has not been addressed from numerical simulations before.

\begin{table*}
\begin{center}
\caption{Description of the simulations}
\begin{tabular}{ccccccc}
\hline\hline
DM profile & r$_{\rm s}$ or r$_{\rm c}$ (pc) & $M_{\rm DM, r200}$ (M$_{\odot}$)  $^a$ & $M_{\rm DM, virial}$ (M$_{\odot}$) $^a$ & $n^{\rm gas}_{\rm c}$ (cm$^{-3}$) $^c$ & SNr/yr & $M^{\rm gas}$/$M^{\rm gas}_{\rm 0}$ \\
\hline
Log & 200 & $1 \times 10^7$ & $1 \times 10^9$ & 1.0 & $1 \times10^{-6}$ & $0.96$ \\
Log & 200 & $1 \times10^7$ & $1 \times 10^9$ & 1.0 & $1 \times10^{-5}$ & $0.58$ \\
Log & 200 & $1 \times10^7$ & $1 \times 10^9$ & 1.0 & $1 \times10^{-4}$ & $0.23$ \\
Log & 200 & $1 \times10^6$ & $1 \times 10^8$ & 1.0 & $1 \times10^{-6}$ & $0.74$ \\
Log & 200 & $1 \times10^6$ & $1 \times 10^8$ & 1.0 & $1 \times10^{-5}$ & $0.28$ \\
Log & 200 & $1 \times10^7$ & $1 \times 10^9$ & 10.0 & $1 \times10^{-5}$ & $0.44$ \\
Log & 200 & $1 \times10^7$ & $1 \times 10^9$ & 10.0 & $1 \times10^{-4}$ & $0.04$ \\
\hline
NFW & 200 & $1 \times10^7$ & $1 \times 10^9$ & 10.0 & $1 \times10^{-4}$ & $0.11$ \\
NFW & 500 & $1 \times10^7$ & $1 \times 10^9$ & 10.0 & $1 \times10^{-4}$ & $0.04$ \\
Log & 200 & $1 \times10^7$ & $1 \times 10^9$ & 10.0 & $1 \times10^{-4}$ & $0.03$ \\
Log & 500 & $1 \times10^7$ & $1 \times 10^9$ & 10.0 & $1 \times10^{-4}$ & $0.40$ \\
\hline\hline
\end{tabular}
\end{center}

{\begin{flushleft}
$^a$ Dark matter mass enclosed at $r=200$, at $z=0$.

$^b$ virial mass enclosed the virial radius, defined where the overdensity is $\rho_{200} = 200 \rho_{\rm crit}$.

$^c$ $n^{\rm gas}_{\rm c}$ is the number density of the gas at the center of the galaxy.
\end{flushleft}
}

\end{table*}

\section{Results} 

Initially, in order to study the energy budget of SNe explosions and the gravitational potential of the dark matter total mass, we run 7 models varying the DM halo masses, for which we used the logarithmic dark matter distribution (Eq.13), and star formation rates as described in Table 1. 

\begin{figure}
	\centering
		\includegraphics[scale=0.4]{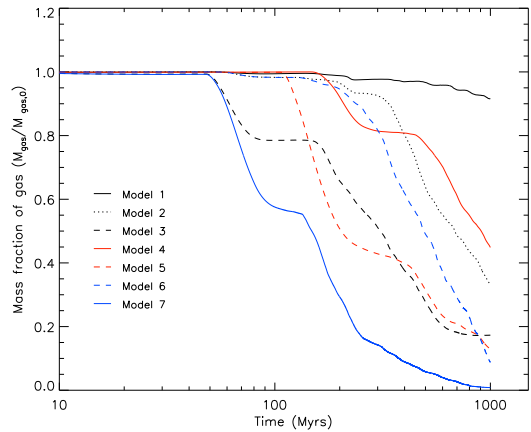}
		\caption {Remaining gas mass normalized by its initial value as a function of time for each of the models described in Table 1.}
	\label{fig2}
\end{figure}

The total baryon mass within the simulated box for each model as a function of time is shown in Figure 2. In the plot, each curve corresponds to the remaining gas mass at a given time, normalized by its initial value. Lower values correspond to a larger mass loss during the simulation, also steeper gradients correspond to increased mass loss rates. We find that all the models considered in this work present, 
though at different levels, mass loss. The correlation between the mass loss and the physical properties of each model is discussed as follows.

\subsection{Mass loss {\it vs} DM halo mass}

In order to understand the role of the dark matter potential in the mass loss 
rate we may compare the results of Models 1, 2, 4, and 5. Models 1 and 4 present 
equal initial physical conditions, except for the dark matter mass. The same 
with models 2 and 5. 

In Model 1, the SNe driven wind is responsible for removing $\sim 4$\% of the gas 
after $500$Myrs, while in Model 4 the total loss is of $\sim 26$\% considering the same 
timescale. After 1Gyr, 
Model 1 shows $\sim 9$\% of mass loss and Model 4 around $\sim 65$\%, as seen in 
Fig. 2. Obviously, the gravitational binding energy here is responsible for this difference, as already noted previously in former numerical simulations (e.g., Mac-Low \& Ferrara 1999). The lower the dark matter total mass the larger the gas mass lost in winds. 
However, as seen in the comparison of these two models, the relationship between 
the binding energy and the mass loss efficiency is not linear. 

The comparison of Models 2 and 5 results in a similar conclusion, but with some 
complications. Model 1 has a large binding energy and a small SN rate that resulted 
in a late and slow wind. Model 2, which is set with a SN rate one order of magnitude 
larger, presents a much stronger wind. Also, the mass loss rate is small up to $t 
\sim 300$Myrs but increases fast after this time, subtly reducing the mass fraction 
of gas. 

This ``two stage" profile is detected in most of the models and are related to 
two processes: i- the dominant mechanism of wind 
acceleration, and ii- the migration of the SNe location to outer radii. 
In the first case, the slow wind is caused by the kinetic pressure of 
the SNe near the center of the galaxy. The abrupt increase of the mass loss process 
is a result of the Rayleigh-Taylor instability (RTI), as will be discussed in the next 
section. In the second scenario, the initial bursts generate a cavity of low density 
gas surrounded by a dense shell. The star formation keeps going at this shell but no 
more energy is injected from stars formed within it. 
This process results in a temporary decrease in the SFR, which halts the galactic wind for a short 
period, until the gas falls back into the central region of the galaxy and feeds star 
formation once again.

After $t = 500$Myrs of the first stars birth, Model 2 presents 
a total loss of $\sim 42$\% of the initial gas mass while Model 5 presents a loss 
of $\sim 72$\%. Again, this difference is related to the gravitational binding energy. 
However, at  $t = 1$Gyr, the total gas mass loss reaches $\sim 67$\% and $\sim 88$\% 
for Models 2 and 5, respectively. The curves shown in Fig. 2 for these two models 
present ``multiple stages", characterized by different slopes, i.e. different mass 
loss rates. Notice that at the end of the simulations the mass loss rate for Model 2 is even 
larger than for Model 5 (the slope in Figure 2 is steeper). This could be explained by a mass loss rate depending on the remaining 
amount of gas in the galaxy. As the system loses gas in earlier epochs the SFR diminishes and 
the galactic wind weakens. However, this is not the case for these two models. 
Here, the RTI is speeding up the mass loss process of Model 2 at later epochs.

\begin{figure*}
	\centering
		\includegraphics[scale=0.23]{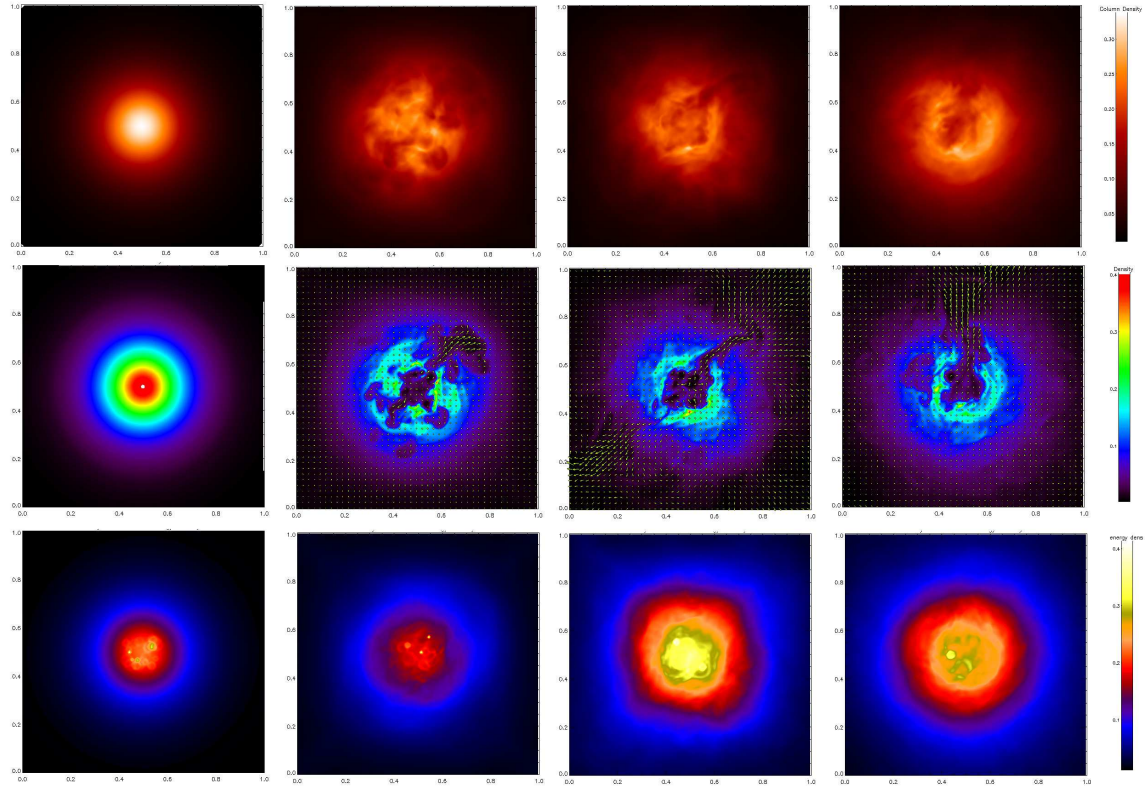}
		\caption {Column density (top), gas density map (center) and thermal energy 
density (bottom) for $t=0, 50, 100$ and $200$Myrs, from left to right, obtained from Model 3.}
	\label{fig3}
\end{figure*}

\subsection{Rayleigh-Taylor instability}

The RTI occurs when a low density gas interpenetrates a denser 
one, as the cavities created by SNe explosions buoyantly rising up through the denser 
ISM. The perturbation grows exponentially (i.e. $\delta v \propto \exp(\gamma t)$) at 
a rate given by (Ristorcelli \& Clark 2004): 

\begin{equation}
\gamma = \sqrt{ g k \frac{\rho_{\rm ISM} - \rho_{\rm cav}}{\rho_{\rm ISM} + \rho_{\rm cav}}},
\end{equation}

\noindent
being $k$ the wavenumber of the perturbation, $g$ the acceleration of gravity and 
$\rho_{\rm cav}$ and $\rho_{\rm ISM}$ the densities of the SNe blown cavity and of the 
ISM, respectively. 

Interestingly, the RTI is more important in galaxies with more massive 
haloes, or in more concentrated haloes. 
This result contradicts the basic assumption derived from Eq. 1, namely the requirement for the kinetic energy to be larger than the gravitational binding energy. The critical point is that 
a homogeneous ISM is implicit in Eq. 1, though in reality SNe explosions create cavities of lower 
density gas that may be buoyantly unstable, which will be stronger in more massive haloes. 

The RTI also plays a major role in Models 6 and 7. These runs were set with the same parameters as 
Models 2 and 3, respectively, except for the larger ISM gas density. As we see in Figure 2, Models 6 
and 7 lost more mass compared to the later two. In terms of energy conservation, for an homogeneous 
ISM, the galactic wind should only depends on the energy injected by SNe (i.e. on the SFR) and 
on the gravitational potential. However, as mentioned before, the RTI growth rate depends on the 
difference between the cavity and ISM densities, i.e. denser ISM result in a faster rise of the 
low density gas. 

It is possible to estimate the timescales for the RTI to efficiently raise the low density gas out 
of the galaxy. If the perturbations are large compared to the dissipation or injection scales (i.e. $h 
> 10 - 20$pc) the instability is assumed to enter in a self-similar growth phase, and the mixing length is then approximatelly equal to: 

\begin{equation}
h(t) \sim 0.05 g t^2 \frac{\rho_{\rm ISM} - \rho_{\rm cav}}{\rho_{\rm ISM} + \rho_{\rm cav}}. 
\end{equation}
 
For $\rho_{\rm cav} = 0.5 \rho_{\rm ISM}$, $h = R_{gal} \simeq 300$pc at $t = \tau_{\rm RT} \sim 100$Myrs for 
$M_{\rm DM, virial}=10^9$M$_{\odot}$, and $\tau_{\rm RT} \sim 290$Myrs for $M_{\rm DM, virial}=10^8$M$_{\odot}$. If 
$\rho_{\rm ISM}/\rho_{\rm cav}$ is ten times larger, as we varied in the simulations, the timescales will be $\sim 2$ 
times shorter.

The buoyant rise of the SNe inflated cavities is clearly seen in Figure 3, where we show the column density (top), central slice of density (middle) and integrated thermal energy density (bottom), at 
different evolutionary stages of the run for Model 3. 

Just few million years after the first starburst SNe start the heating and stirring of the ISM, as 
shown in the left panel ($t=0$) of the thermal energy density projection. Each bright spot represents 
a SN explosion. Its further evolution, with cooling taken into account, results in stalled dense 
shells surrounding low density cavities. These cavities are seen in the density map (middle row) 
for $t=50$Myrs, as well as for the column density (top row). When several explosions occur close 
to each other a large cavity is created. This low density bubble becomes convective unstable and 
rises up, as seen at the density maps and column density projections between $t=50$ and $t=100$Myrs. 
The instability is then responsible for the creation of a tunnel through which gas may scape. 

Regarding the gas flows, the enhanced gas velocity through these tunnels is also seen in 
Figure 3 as normalized vectors over-plotted to the density maps. For Model 3, it is clear 
that the mass loss is not spherically symmetric. However, the tunnels created by the RTI are 
short lived. Once the potential energy of the cavities are released, the flow that continues 
to rise through them do not present pressure large enough to prevent the ISM to collapse it. 
In Figure 3, it is clear the destruction of the diagonal filament between $t=100$ and $t=200$Myrs. 
In the mean time an another bubble is created and rose up in the vertical direction, as seen at 
$t=200$Myrs. 

The cyclic process of buoyantly unstable cavities is important to reactivate star 
formation at the central region of the galaxy. When a large cavity is created the star 
formation cease due to the decrease in density below the critical threshold. At this time, the 
star formation migrated to the surrounding denser shell. This is clearly seen in the 
integrated energy density maps of Figure 3. The bright spots, representing the 
SN explosions, are concentrated in the center of the galaxy at $t \le 50$ but 
moves to larger radii at $t \ge 100$. When the cavity moves upward due to the 
RTI, the ISM denser material occupies the center of the galaxy once again, bringing the local 
density above the critical value. 

The competition between the RTI and kinetic pressure may be estimated by comparing two timescales. 
One is the RTI timescale described above, the other is the timescale for the SNe to release 
energy into the ISM:

\begin{equation}
\tau_{\rm kin} \sim \frac{E_{\rm b}}{\eta \epsilon_{\rm SN} R_{\rm SN}}.
\end{equation}

\noindent
If $\tau_{\rm kin} < \tau_{\rm RT}$, the kinetic energy released by the SNe 
will result in galactic winds before the RTI is able to rise the buoyancy 
cavities out of the galaxy. In other words, for a typical dSph, the RTI 
will play a major role in releasing part of the energy injected by SNe if:

\begin{equation}
R_{\rm SN} < \frac{7 \times 10^{-7}}{\eta} \left( \frac{M_{\rm DM, r200}}{10^7 M_{\odot}} 
\right)^{3/2} \left( \frac{\rho_{\rm ISM} - \rho_{\rm cav}}{\rho_{\rm ISM} + \rho_{\rm cav}} 
\right)^{1/2} {\rm SN\ yr^{-1}}.
\end{equation}

The fraction of the mass loss due to the RT instability $\dot{M}_{\rm RT}/\dot{M}_{\rm tot}$ may then be estimated. Numerically, we flag each buoyantly unstable cell of the cube and track its dynamical evolution outwards (basically the density $\rho_{\rm RT}$ and velocity field ${\bf v}_{\rm RT}$.) 

\begin{equation}
\frac{\dot{M}_{\rm RT}}{\dot{M}_{\rm tot}} = \frac{\int{\rho_{\rm RT} (t) {\bf v}_{\rm RT} (t) \cdot {\bf dA}}}{\int{\rho_{\rm gas} (t) {\bf v}_{\rm gas} (t) \cdot {\bf dA}}},
\end{equation}

\noindent
where $A$ is defined as the outer boundary of the simulated domain.

In Figure 4 we show slices of the flagged unstable cells, overplotted with the corresponding velocity vectors, for Models 1, 2, 5 and 6, at $t=50$Myrs. The color scheme corresponds to the local density of the unstable cells.

\begin{figure*}
	\centering
		\includegraphics[scale=0.22]{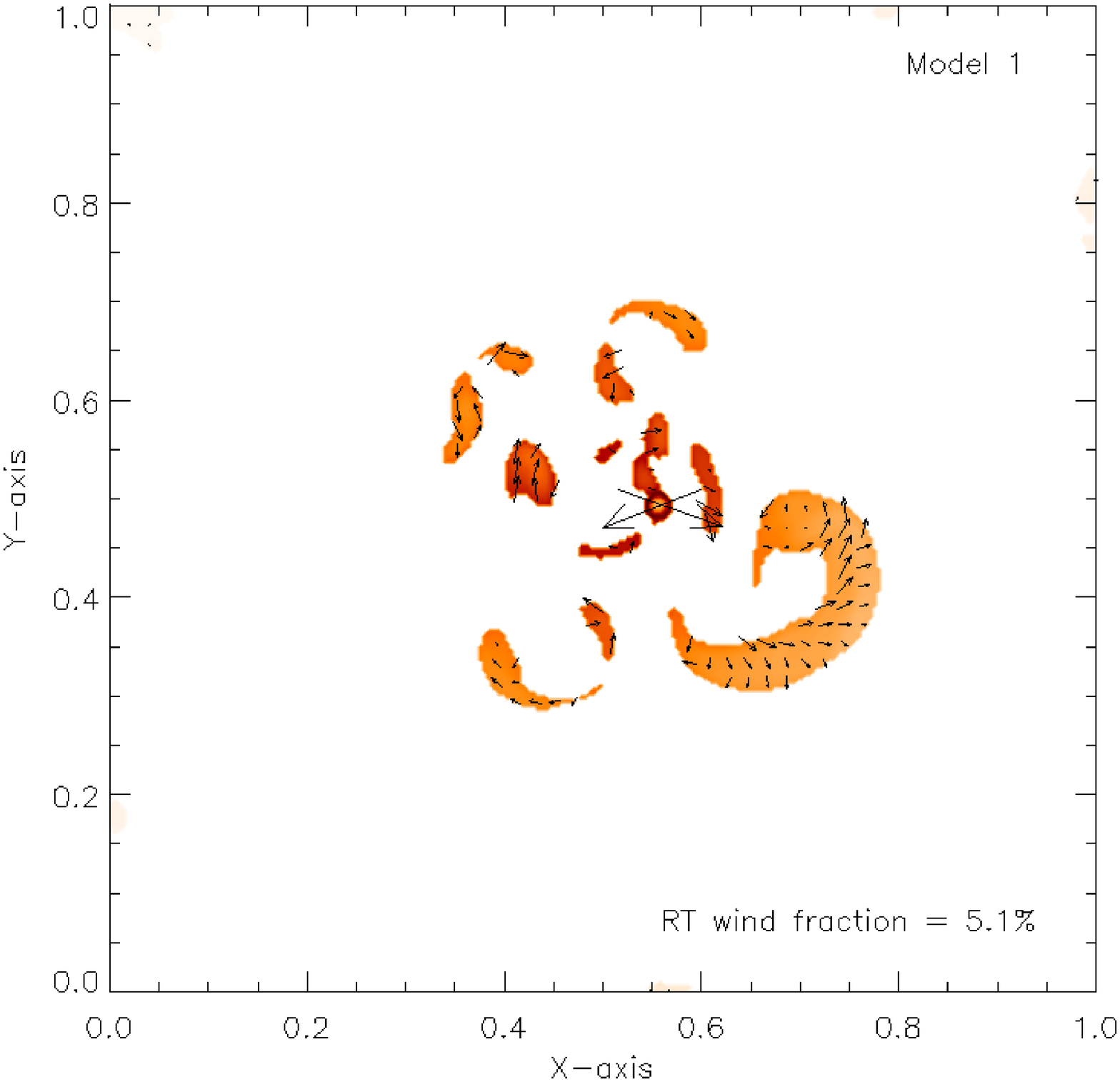}
		\includegraphics[scale=0.22]{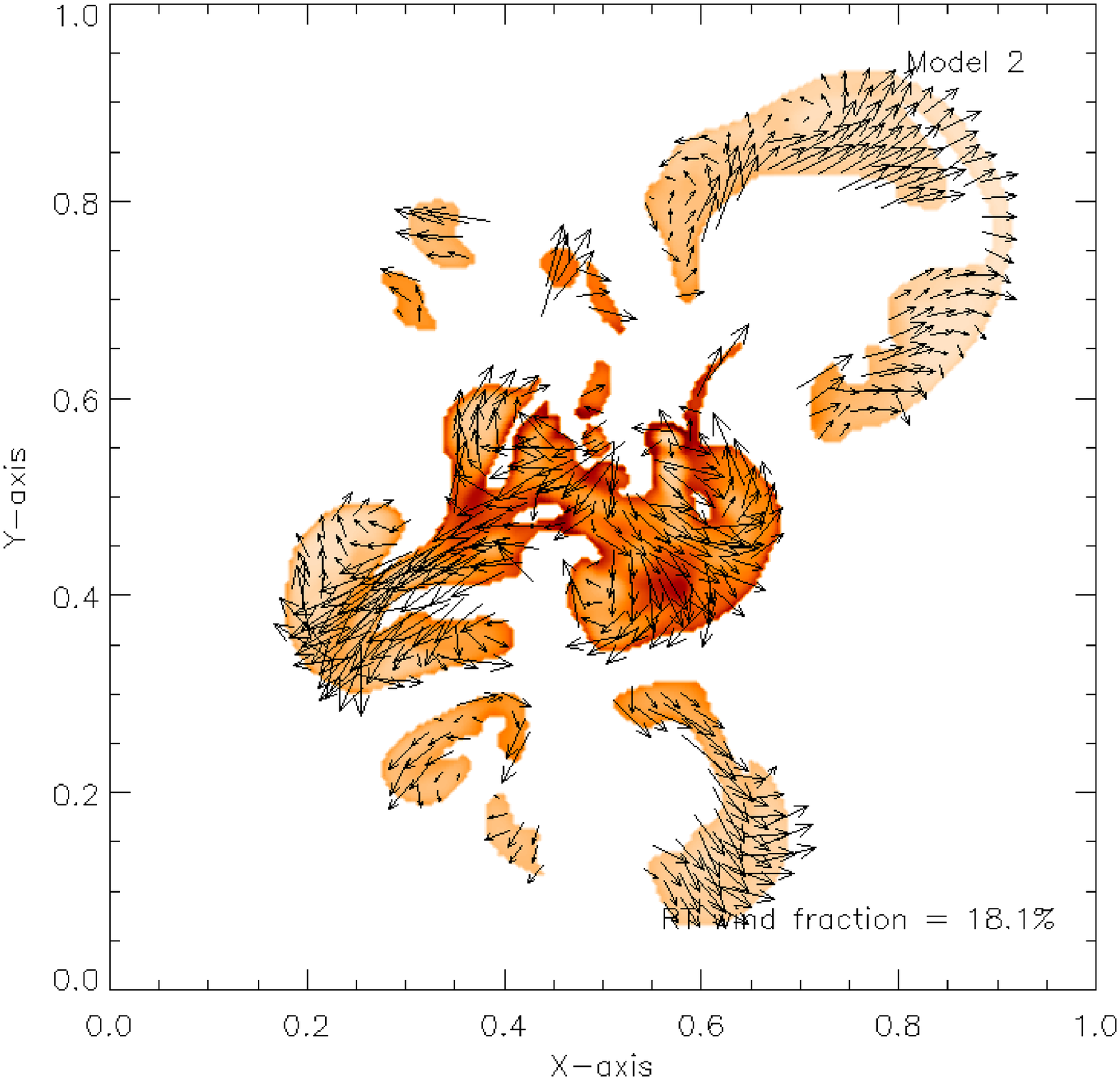} \\
		\includegraphics[scale=0.22]{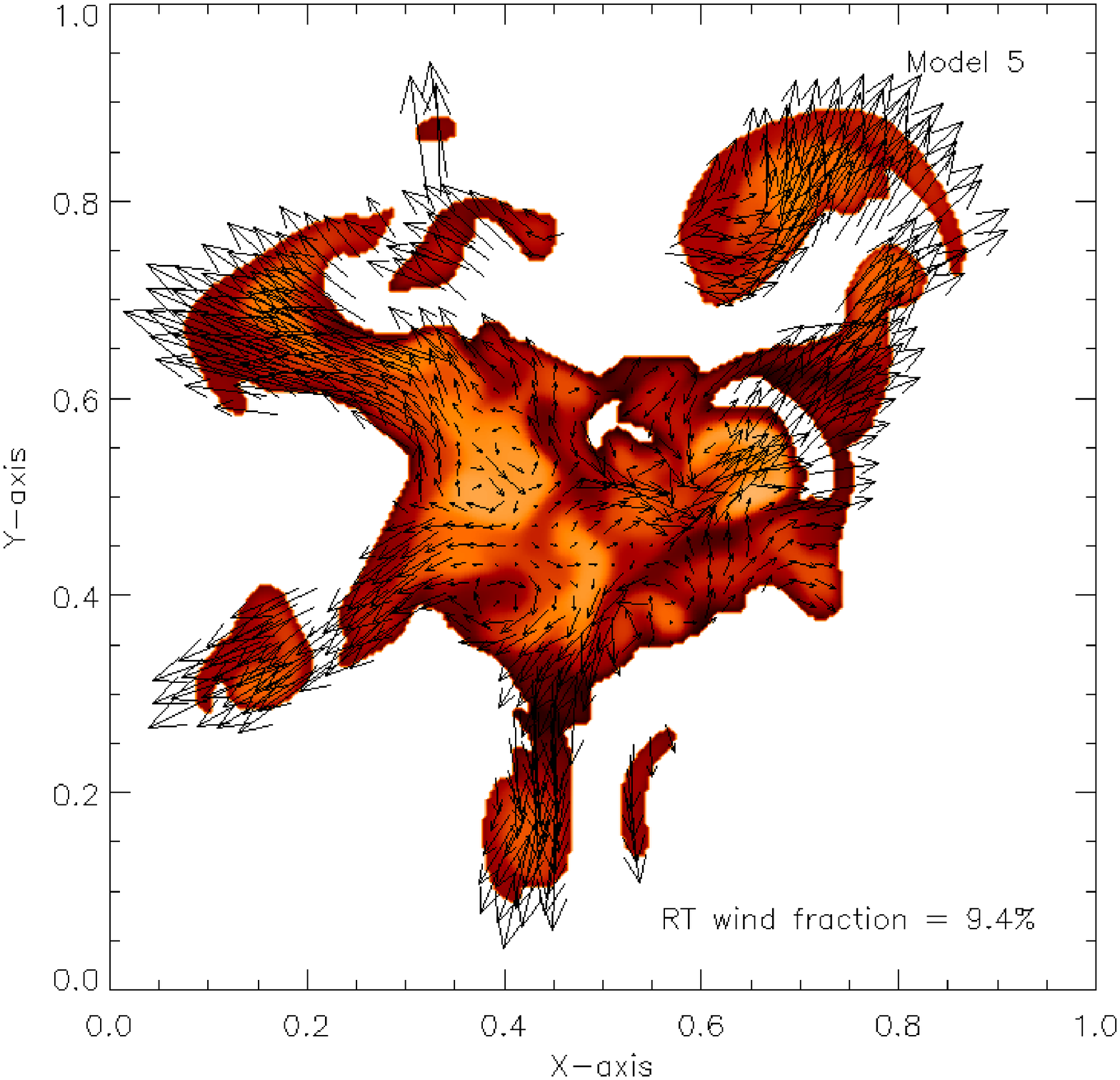}
		\includegraphics[scale=0.22]{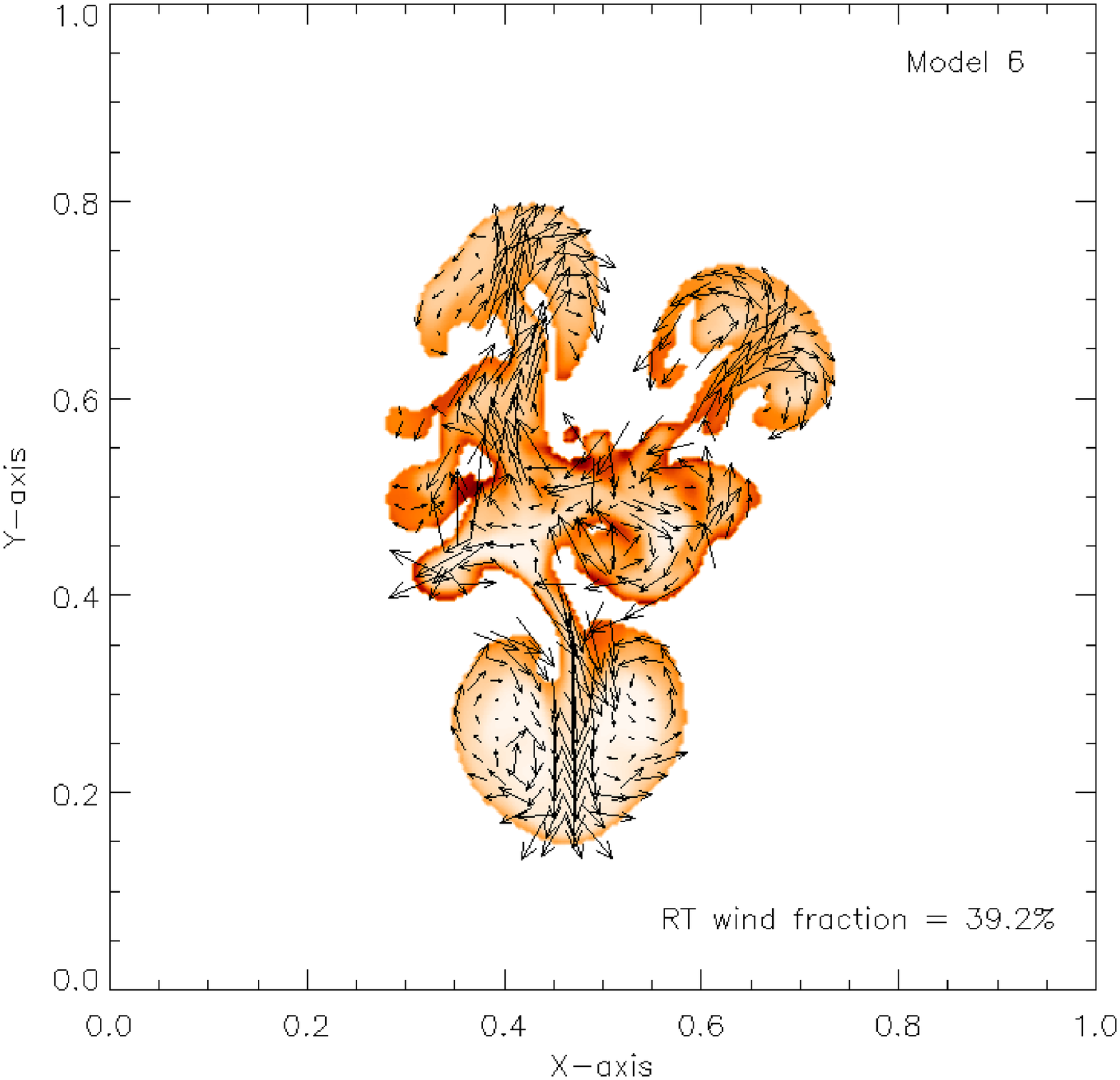}
		\caption {Maps of density overplotted to with velocity vectors of Rayleigh-Taylor unstable regions in Models 1, 2, 5 and 6, at $t=50$Myrs. The fraction of the Rayleigh-Taylor excited galactic wind at the given time, relative to the total, is also indicated.}
	\label{fig4}
\end{figure*}

The relative mass loss rate due to the RT instability is also presented in the plots. Obviously, for Model 1, where the rate of explosions is low, it is more difficult for the interacting SNe to generate large bubbles. The wind obtained in this model is basically wave-driven. Model 2, with 10 times more explosions, result in an increased RT-driven wind. However, the dark matter mass and the local density also play a role. Model 5, with a dark matter mass 10 times smaller than in Model 2, result in lower RT-driven mass loss rate. In this case, the explosions push the gas from the central region outwards resulting in increase mass loss rates, but dominated by the direct transfer of momemtum from the SNe to the gas. Also, as seen for Model 6, the increased local gas density is less influenced by the SNe directly, but increases the RT-instability, resulting in a relative importance of $\sim 40$\% in the mass loss rate. The muchroom like structures are perfectly recognized in this model.
It is also seen from Figure 4 that the rising bubbles are modified by Kelvin-Helmholtz and Karman vortex instabilities, which destroy the spherical morphology of the cavities, excite turbulence behind them and enhance the gas mixing with the colder external gas.

We obtained a mass loss rate contribution of $\sim 5 - 40$\% due to the RT instability for all models. This effect though is strongly at $t > 200$Myrs. Once the quasi-steady wind is developed, it becomes more difficult to precisely estimate the fraction of the mass loss due to the RT instability. Basically, most of the gas in the center of the cube is heated and part of it also gain momemtum from the SNe explosion. In this sense we expect a transition in the chemical enrichment of the intergalactic medium due to the galactic winds.

\subsection{Mass loss {\it vs} SNe rate}

Despite of the different particular mechanisms of galactic wind acceleration 
(e.g. kinetic pressure and convective instability), the main source of energy and momentum for 
all these is still stellar feedback. It is important then to study the relationship 
between the mass loss rates obtained from the simulations and the preset rate of SNe ($R_{\rm SN}$).

The SNe rate is the only difference in the initial setup of the Models 1, 2 and 3. The remaining baryon mass for each model at $t=500$Myrs is shown in Table 1. Model 3 presents the lowest value, which is related to its larger SFR, and $R_{\rm SN}$ as a consequence. Model 1, on the other hand, presents a small mass loss, of only 4\%, compared to 42\% for Model 2, and 77\% for Model 3. At $t=1$Gyrs, these values are $\sim 9$\%, 70\% and 85\%, respectively. The correlation between  $R_{\rm SN}$ and mass loss rate is also clear, as shown in Figure 5. This is also true for the other models.

As seen in Figure 5, the mass loss process is not stationary, varying in short timescales, with the presence of peaks of fast mass loss and periods of weak, or even absent winds. As example, Model 3 presents two events of increased mass loss rates, with $\dot{M} \sim 0.9$\% Myr$^{-1}$ at $t \sim 60$Myrs and $\dot{M} \sim 0.35$\% Myr$^{-1}$ at $t \sim 150$Myrs, decreasing slowly with time afterwards, while Model 2 present a very similar profile, with a single peak ($\dot{M} \sim 0.3$\% Myr$^{-1}$), but much later around $t \sim 400$Myrs. At this stage, the mass loss rate of Model 2 is even larger than in Model 3.

For Models 4 and 5, and Models 6 and 7, the correlation of mass loss rates with SFR is even clearer. The peak in mass loss rate for Model 4 is two times smaller, and occur $\sim 70$Myrs later than for Model 5. Model 6 presents a peak in mass loss rate almost 5 times smaller, and delayed in $\sim 300$Myrs compared to Model 7. For this case in particular the RTI plays a major role on this difference.

\begin{figure}
	\centering
		\includegraphics[scale=0.45]{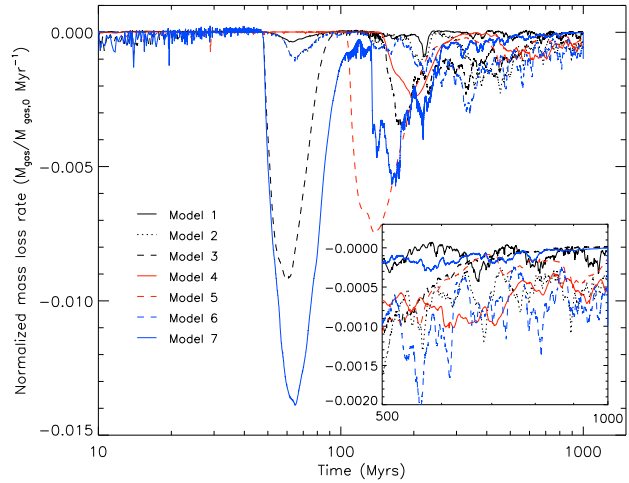}
		\caption {Mass loss rate as a function of time for each of the models described in Table 1.}
	\label{fig5}
\end{figure}

It is worth noticing that the duration of the strong winds is also dependent on $R_{\rm SN}$. Basically this is due to the loss of gas, which in turn supplies further star formation. Once the galaxy presents a strong wind most of volume will not present ideal physical conditions for new starbursts. In some cases, gas will fall back to the core of the galaxy, triggering a new sequence of SNe, as seen in Models 3 and 7.

\subsection{Dark matter density profile}

In order to test the role of different dark matter distributions in the galactic wind acceleration and instabilities, we additional run 4 models, considering the two different density distributions (namely the NFW and logarithmic), for two different concentrations radii, as described in Table 1. 

The mass distribution of the modelled dark matter radial distributions are given in Figure 6 (upper panel). The plot refers to the total dark mass enclosed as a function of the radius. Notice that we have assumed all distributions to give same values at $r=200$pc. This follows the result given in Walker (2012) (see its Figure 18) showing that fits to the velocity dispersion of stellar population in dSph galaxies predict approximately the same enclosed mass near the halflight radius, regardless the dark matter distribution that is adopted. The time evolution of the total gas mass in each case is given in the bottom panel of Figure 6. We find that all models give strong winds and large mass loss in the timescales of $t \rightarrow 1$Gyr. Model 11, with a logarithmic DM density profile with concentration radius $r_{\rm c} =500$pc presents the largest delay in the mass loss rate, compared to the other 3 models that present very similar mass loss rates as function of time. The reason for this difference is the lower gravity acceleration at lower radii for Model 11. This may be related to a stronger RT-instability occuring at models where the dark matter potential is larger at smaller radii. In this sense, models with NFW dark matter distributions would tend to show stronger RT instabilities.

\begin{figure}
	\centering
		\includegraphics[scale=0.3]{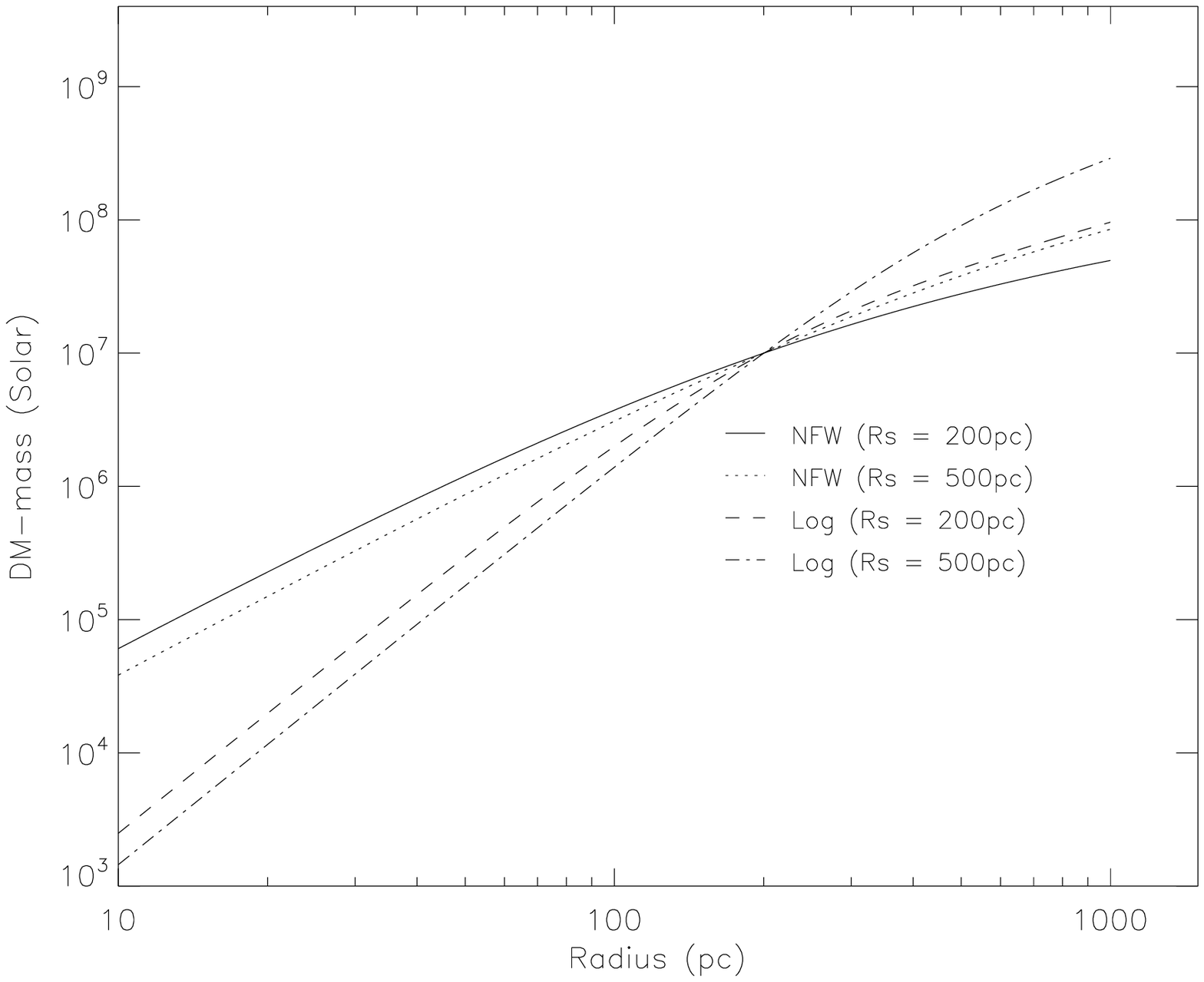}
		\includegraphics[scale=0.3]{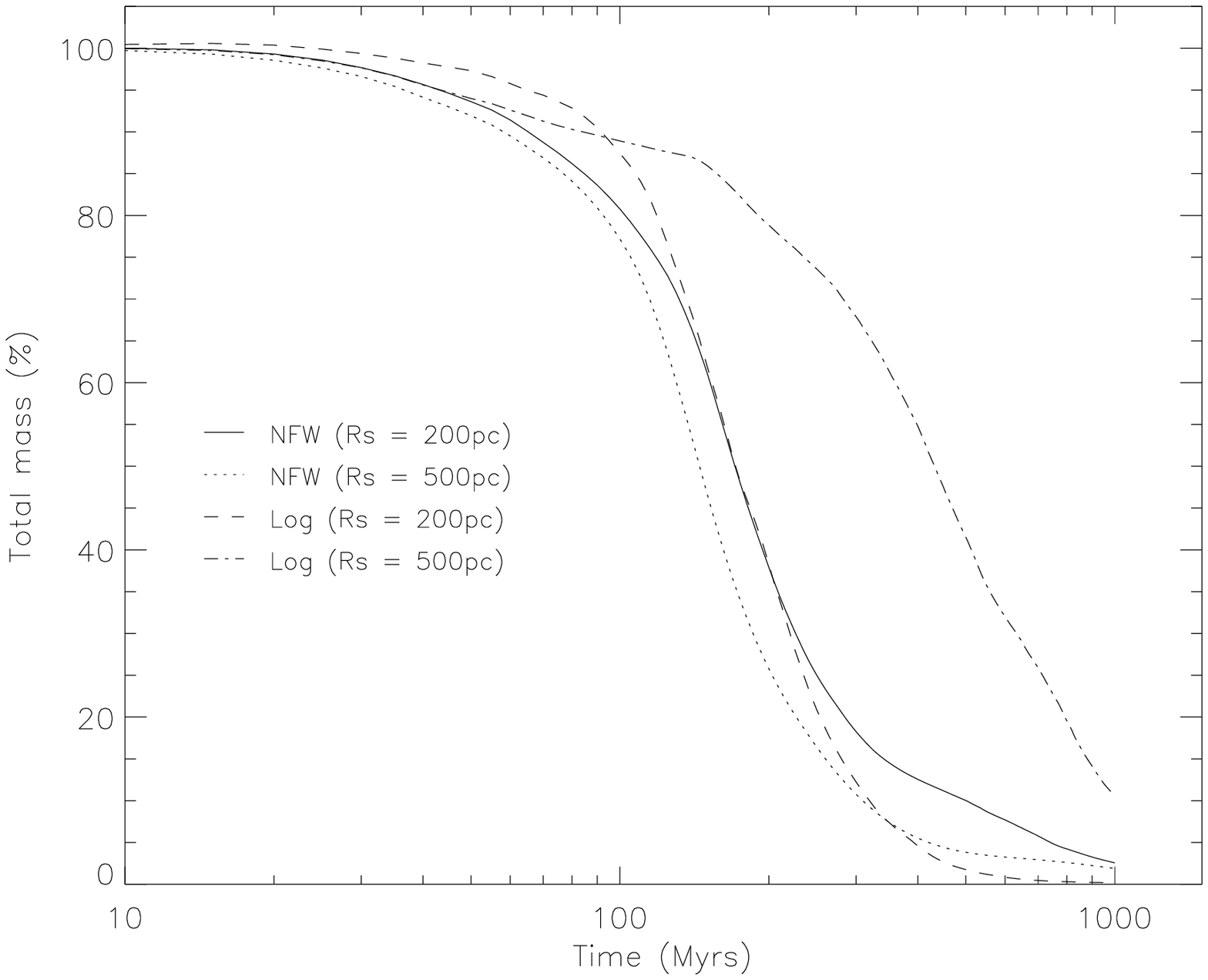}
		\caption {Up: total enclosed dark matter mass as function of radius for Models 8 - 11 (see Table 1). Bottom: total remaining gas mass as a function of time.}
	\label{fig6}
\end{figure}

\section{Discussion}

The numerical simulations provided in this work show that galactic winds may have occurred in the early stages of most of the dwarf galaxies. Several previous works studied the same problem, considering similar schemes, namely considering SNe as the main source of energy to trigger the galactic wind. Despite of the numerical resolution, which is responsible for relatively important changes in the obtained results, the major difference between these and the present work relies on the treatment of the SNe as a random event, both in time and in space. Also, the treatment of the radiative cooling homogeneously in the computational grid is also different to the prescriptions used in the past.

\subsection{Comparison to previous works}

Marcolini et al. (2006) studied the SNe feedback in the chemical evolution of the particular case of the  Draco dwarf spheroidal galaxy using a similar scheme. These authors computed few 3D numerical simulations with a maximum spatial resolution of 13pc per cell, compared to the 1.95pc per cell in our case. In their scheme, the SNe are supposed to occur in a sequence of instantaneous and identical bursts. In contrast to our model, the SN events were random in space but not random in time. According to the authors, the simultaneous bursts resulted in a total energy released by the SNe orders of magnitude
larger than the binding energy of the ISM. Even though, the galactic wind produced was negligible. This was due to the large efficiency of the radiative cooling.
We note that, in their model, the assumed star formation history resulted in a SNe rate below 
$10^{-6}$SN yr$^{-1}$ for the whole simulation (up to $t=3$Gyr). This result 
is in perfect agreement with our simulations. 

As in their specific case, we found that even when the SNe result in a larger energy released compared to the gravitational binding energy, the radiative cooling is efficient in removing most of this energy from the system. The consequence is a much lower net kinetic/thermal pressure of the ISM, which inhibits the mass loss. Besides, we provide a more detailed study of the physical parameters and show that, despite of the efficient cooling of the SNe heated gas, the galactic wind may be important when restricted conditions are fulfilled.

Fragile et al. (2003) also studied the dynamical evolution of the gas component during the early stages of dwarf galaxies. They have already recognized that the results are sensitive to numerical overcooling of the shocked gas, which leads to an underestimation of the effects of the supernovae. In order to prevent this overcooling they did not compute cooling within the SN-ISM shock until it has already evolved the Sedov phase. Despite of the unrealistic treatment of the physics in their scheme, due to the computational limitation at that time, they have realized that, in the absence of cooling, the hot and enriched gas from the supernovae would eventually leave the galaxy because of their buoyancy. They also pointed out that the numerical overcooling would prevent this effect. 

In our work, we calculated the radiative cooling of the models using a high-order numerical scheme and finer grid resolution compared to the previous works. This allowed a self-consistent treatment of the dynamical evolution of individual supernovae. We also find that radiative cooling plays a major role in the dynamical evolution of the ISM gas. The cooling timescale $\tau_{\rm cool} \simeq kT/n\Lambda$, gives $\tau_{\rm cool} > \tau_{\rm RT}$ for the bubbles, but is very short (compared to the dynamical timescales) for the typical ISM. It means that the thermal energy of the ISM gas is quickly lost by radiation, long before the thermal/kinetic pressure is able to drive the galactic winds. However, the low density cavities present longer cooling timescales, which result in efficient RTI driven outflows.

The star formation history of dSph galaxies strongly depends on the properties of their winds. We have shown that, depending on the dark matter profile and SN rate, the thermal and kinetic pressures at the center of the galaxy push the gas outwards. The result is a low density core gas, and the star formation may be quenched. Once the central region cools and kinetic pressure is dissipated, the envelope gas falls back into the galaxy and new stars form. Notice that cyclic starbursts due to episodic galactic winds have also been obtained in SPH numerical simulations (e.g. Stinson et al. 2007). Numerical tests though show that the Rayleigh�Taylor instability is poorly resolved by SPH techniques. This because SPH introduces spurious pressure forces bacause of the smoothing kernel radius, over which interactions are severely damped (Agertz et al. 2007). In this sense, it is clear that the RT instability is possibly not the dominant mechanism in quenching star formation at the center of the galaxy, nor in triggering the galactic wind. However, it may be important in fast delivering metal enriched gas to the intergalactic medium - long before the diffusion of the elements at local ISM is possible -, and on the release of thermal energy from the core.

\subsection{Winds and the chemical evolution of dwarf galaxies}

Mass loss is claimed to be one of the main mechanisms that controlls the chemical evolution of dSph galaxies. The removal of a considerable fraction of the gas content of the galaxy will affect directly the star formation and also the production of new chemical elements and the enrichment of the ISM. For instance, Recchi, Matteucci \& D`Ercole (2001) studied the 2D-chemodynamical evolution of the gas in a dwarf irregular galaxy (dIrr) considering the effects of SNe driven winds. The authors concluded that selective galactic winds, in terms of elements that are removed, can explain most of the observed abundances in these objects. This could possible be applicable to the early stages of dSph galaxies as well. If this is the case several observational constraints will have their pattern defined by the mass loss rate. The assumption of a galactic wind occurring during the evolution of the galaxy can explain, for instance, the mass-metallicity and mass-luminosity relations, as well as the relation between [O/H] and mean velocity dispersions, observed in dSph galaxies. Richer, McCall \& Stasinska (1998) suggested that there is a correlation between [O/H] and mean velocity dispersions in several types of dynamically hot galaxies that can be explained naturally if the chemical evolution proceeded until the energy input from SNe gave rise to a galactic wind. A similar scenario was presented by Tamura, Hirashita \& Takeuchi (2001), where they explain the mass-metallicity relation as due to the SF proceeding at a very low rate until a galactic wind develops and expels the gas out of the system. 

From the theoretical point of view, galactic winds were also suggested as a necessary mechanism to explain some chemical properties of local dSph galaxies, even though there is a controversy regarding the rate of the gas loss. Kirby et al. (2011) claimed that the observed low metallicities cannot be fitted by a closed box model of chemical evolution (given the long timescales for the SF in such galaxies) implying that the removal of gas is required to avoid an abnormal increase in metallicity. Besides, the lowest values of the [alpha/Fe], the trends of the neutron capture elements and the shape of the stellar metallicity distributions can be very well reproduced by models of chemical evolution that adopted SNe triggered galactic winds, with a high efficiency which is adjusted by hand to fit the data (Lanfranchi and Matteucci 2004, 2007). The removal of a large fraction of the gas will decrease substantially the SFR, almost ceasing the formation of new stars and the injection of oxygen and r-process elements in the ISM. Iron and s-process elements, on the other hand, are produced and injected in the medium in a much longer timescale (up to some Gyr), giving rise to low abundance ratios after the wind starts. The decrease in the SFR will also prevent the formation of stars with high metallicities, keeping the mean metallicity of these galaxies low with a stellar metallicity distribution peaked at [Fe/H] below -1.4 dex.

Different kinds of simulations, either hydrodynamical (Marcolini et al. 2006, 2008) or with SPH codes (Revaz et al. 2009, Revaz \& Jablonka 2012), reach, however, other conclusions regarding the efficiency of gas removal. In the SPH simulations, normally, the hydrodynamics of the gas is not treated in detail and the efficiency with which thermal energy is converted into kinetic energy is a free parameter adjusted ad-hoc. Revaz \& Jablonka (2012), in their analysis of the dynamical and chemical evolution of dSph galaxies, adopted a low value for the star formation feedback efficiency in their simulations to fit the final predicted metallicity to the observed values. If the feedback efficiency is too high, the metallicity would be below what is observed. In fact, these authors claim that the most sensitive parameter in their code is this efficiency and that the low values adopted (between 0.03 and 0.05) imply that strong winds are not compatible with the observations. The main causes for the low values for the feedback efficiency are, however, not treated. In the simulations of Marcolini et al. (2006, 2008), on the other hand, the dynamics of the hot gas was taken into account carefully, but the galactic wind produced was very weak and capable of removing only a small fraction of the gas. The main reason behind that result, is that the adopted efficiency of the radiative cooling was large. In none of these cases (nor in any other study) the effects of the instabilities in the ISM caused by the low density gas that interpenetrates a denser region due to the SNe explosion is considered. As previously mentioned, the occourence of RTI can alter significantly the scenario for the occurence of the galactic wind, in respect of both the epoch when the gas starts to be removed and the rate at which the gas is lost.

The epoch when the gas starts to be removed and the rate of this process are crucial in the models of chemical evolution. The rate of the wind is, in general, a free parameter adjusted just to fit observational constraints, and the time when it begins depends on the energetics of the galaxy: it starts when the kinetic energy of the gas exceeds the binding energy of the galaxy. The first term depends on the SNe rate and the last one on the mass of the dark halo (Bradamante et al. 1998, Lanfranchi \& Matteucci 2003). The relation between these terms and the mass loss is not evident, as shown by the hydrodynamical simulations presented in this work. As discussed in the previous sections, gas can be removed from the galaxy even if the kinetic energy is not larger than the binding energy, mostly due to the RTI. The cavities created by SNe explosion carry gas out of the galaxy even in more massive haloes. In fact, this effect can be stronger in these systems. Besides that, the removal of gas is not uniform, as seen in Figure 2. The ``two stage" profile of the mass loss can affect directly the SFR and the enrichment of the medium. Following the oscillation of the SFR, the enrichment of the medium will also vary in time, leading perhaps to a spread in observed abundances. What is similar in the simulations and chemical evolution models is the dependence of the total amount of lost mass on the dark matter halo mass and in the SNe rate. Galaxies with more dark matter tend to present also larger masses for the stellar component, whereas galaxies with higher SNe rate present smaller gas mass at the end of the simulations. Variations in these two quantities affect the amount of gas mass loss and the epoch when gas starts to be removed, giving rise to different chemical enrichment histories (see also Revaz \& Jablonka 2012). Obviously, the details of the enrichment history due to the mass loss as predicted by the simulations can only be fully understood by adopting chemical evolution models that take into account the results presented here.

A major contrast between a RTI-dominated and a wave/turbulent scenario is the origin of the gas that is being removed out of the galaxy. The kinetic pressure tends to push the ISM as a whole upwards, thus removing, first and mostly, the low metallicity fraction of the gas. The RTI, on the other hand, is responsible for the rise of the metal-enriched hot bubbles of gas, resulting in a selective wind. In this case, a large fration of the heavy elements would be ejected out of the galaxy to the intergalactic medium, while the low metallicity ISM would remain. This is in special agreement with both the observations and the predictions of semi-analytical models of galactic chemical evolution for the dSph. This process may also have important cosmological impact since dwarf galaxies are believed to be formed earlier than the more massive ones.

\section{Conclusions}

In this work we presented a number of hydrodynamical numerical simulations of the time evolution of dSph gas component, studying its dependence on the star formation history and the dark matter mass. 

In agreement with previous works, we found a strict, but not trivial, dependence between the mass loss rate and both the star formation rate and the gravitational binding energy of the system. The galactic winds are easily triggered in most of the galaxies, except for those with very little supernova occurence rates ($R_{\rm SN} < 10^{-6}$yr$^{-1}$).

As main results, we conclude that:

\begin{itemize}

\item the complexity of these dependences arise from the different mechanisms that may trigger the galactic wind, e.g. the kinetic pressure and the RTI. More massive haloes tend to inhibit the formation of kinetic/thermal pressure driven winds, but on the other hand accelerate the rise of the buoyant unstable cavities of hot gas.

\item as showed in previous works, the radiative cooling may reduce the efficiency of the SNe in generating galactic winds. However, we found that the RTI may still occur due to the longer cooling timescales of the cavities. 

\item galactic winds may be selective in terms of the elements ejected to the intergalactic medium, depending on the dominant acceleration process. 

This is particularly important in the sense that the dark matter distribution of dSph galaxies may severely change during their evolution. As a consequence, the main wind acceleration mechanism, and the type of chemical elements preferred to be driven out of the galaxy, will be also different.

From the timescales derived in this work it is possible that the winds of dSph have enriched the intergalactic medium before the massive galaxies have been formed.

\end{itemize}

With more computational resources, we plan to extend this work in the near future by 
running more simulations covering with better resolution the range of the important physical 
parameters, such as the star formation rate and the dark matter halo mass. Also, it will be 
interesting to study the difference between kinetic/thermal and RTI-driven winds in terms of 
a multi-fluid hydrodynamical code in order to follow more consistently the dynamical evolution of 
the low and high metallicity fractions of the ISM.

\section*{Acknowledgments}

LOR thanks INCT-A/CAPES (573648/2008-5) for financial support. DFG thanks CNPq (no. 300382/2008-1) and FAPESP (no. 2011/12909-8) for financial support. GAL thanks CNPq (no. 302112/2009-0).

\end{document}